\documentclass[preprint,12pt]{elsarticle}

\usepackage{duckuments}
\usepackage{subcaption}
\usepackage{hyperref}
\usepackage[mode=buildnew]{standalone}
\usepackage[table,xcdraw]{xcolor}

\usepackage{pgfplots}
\usepackage{tikz}
\usetikzlibrary{patterns}
\usepgfplotslibrary{groupplots}
\usepgfplotslibrary{statistics}
\pgfplotsset{compat=1.16}
\usepgfplotslibrary{statistics}
\pgfplotsset{
/pgfplots/boxplot legend/.style={
   legend image code/.code={
     \draw [|-|,##1] (0,1.5mm) --
         node[rectangle,minimum size=2mm,draw,fill,fill opacity=1,##1]{}
        (0,5mm);
    }
  }
}

\AtBeginDocument{%
  \providecommand\BibTeX{{%
    \normalfont B\kern-0.5em{\scshape i\kern-0.25em b}\kern-0.8em\TeX}}}

\usepackage{amssymb}
\usepackage{amsmath}

\journal{Journal of Parallel and Distributed Computing }

\begin{document}

\begin{frontmatter}



\title{To Repair or Not to Repair:\\Assessing Fault Resilience in MPI Stencil Applications} 

\cortext[cor1]{Corresponding author - email:\href{mailto:roberto.rocco@polimi.it}{roberto.rocco@polimi.it}}

\author[poli]{Roberto Rocco\corref{cor1}}
\ead{roberto.rocco@polimi.it}
\author[e4]{Elisabetta Boella}
\ead{elisabetta.boella@e4company.com}
\author[e4]{Daniele Gregori}
\ead{daniele.gregori@e4company.com}
\author[poli]{Gianluca Palermo}
\ead{gianluca.palermo@polimi.it}

\affiliation[poli]{
    organization={DEIB - Politecnico di Milano},
    addressline={Via Giuseppe Ponzio, 34},
    city={Milan},
    country={Italy}
}

\affiliation[e4]{
    organization={E4 Computer Engineering Spa},
    addressline={Viale Martiri della Liberta', 66},
    city={Scandiano (RE)},
    country={Italy}
}


\begin{abstract}
With the increasing size of HPC computations, faults are becoming more and more relevant in the HPC field. The MPI standard does not define the application behaviour after a fault, leaving the burden of fault management to the user, who usually resorts to checkpoint and restart mechanisms. This trend is especially true in stencil applications, as their regular pattern simplifies the selection of checkpoint locations. However, checkpoint and restart mechanisms introduce non-negligible overhead, disk load, and scalability concerns. In this paper, we show an alternative through fault resilience, enabled by the features provided by the User Level Fault Mitigation extension and shipped within the Legio fault resilience framework. Through fault resilience, we continue executing only the non-failed processes, thus sacrificing result accuracy for faster fault recovery. Our experiments on a specimen stencil application show that, despite the fault impact visible in the result, we produced meaningful values usable for scientific research, proving the possibilities of a fault resilience approach in a stencil scenario.
\end{abstract}





\end{frontmatter}



\section{Introduction}

High-Performance Computing (HPC) has just reached the exascale era with the advent of the Frontier supercomputer. Reaching the ExaFLOPS proved a tough challenge, as many overlooked issues emerged. One of these issues is fault presence: while the components used in the cluster should be tested and reliable, the large amount of them makes the probability of encountering a fault non-negligible~\cite{hochschild2021cores, dixit2021silent}. If we want to move past the computation capabilities of Frontier, this probability will grow even more, introducing the need to handle faults within executions. Unfortunately, HPC tools do not always feature fault management functionalities: the Message Passing Interface (MPI)~\cite{clarke1994mpi}, the de-facto standard for inter-process communication, is a striking example of this trend since it does not define the behaviour after fault insurgence.

The absence of fault management functionalities becomes even more problematic when considering the amount of energy consumed by HPC executions. HPC clusters' power consumption has already reached megawatts magnitude, values comparable with those of a small town\footnote{\url{https://top500.org/lists/top500/2024/06/}}. With such an energy impact, it becomes mandatory to get the most out of executions and embrace sustainability principles as much as possible. Among sustainability principles, the ability to repair is a cornerstone since it extends the use life of a good past the insurgence of faults. We must pursue repairability for HPC executions, and we are getting it with efforts like the User Level Fault Mitigation (ULFM) extension~\cite{bland2013post} and the Reinit proposal~\cite{laguna2016evaluating}. Those efforts give the user tools to fix the MPI structure affected by faults or reinitialise the entire MPI layer, respectively. Using ULFM or Reinit, users can repair their executions, reaching the end even in the presence of faults.

While ULFM and Reinit provide functionalities to repair the damage caused by faults, the operation has consequences. Repairing the execution requires time (thus energy), as processes must reconstruct the data damaged or lost. In practice, this usually happens through Checkpoint and Restart (C/R)~\cite{hargrove2006berkeley,ansel2009dmtcp,reber2012criu,garg2019mana}, but it introduces overhead in terms of time and disk load, even in the absence of faults. In the literature, many efforts explored alternatives or enhancements to C/R~\cite{dichev2018energy,losada2019local,filiposka2019multidimensional}, but the approach that introduces the least overhead is graceful degradation~\cite{ashraf2018shrink, rocco2021legio}, where we sacrifice result accuracy for faster recovery. Graceful degradation is powerful and effective when we deal with loosely coupled data, as the loss of a part does not affect the rest of the computations. On the other hand, its use in execution contexts where processes interact a lot will eventually spread the effect of the fault, compromising the result. Despite this issue, if the fault does not have time to spread or does not spread fast enough, we may still get a usable result using a graceful degradation approach in a tightly coupled scenario.


In this effort, we want to focus on this latter possibility, evaluating how and when a graceful degradation approach can outscore a complete repair. We focus our analysis on stencil applications, where processes communicate with their topological neighbours in a regular pattern at each iteration, causing faults to spread over the execution. We analyse the Legio fault resilience framework~\cite{rocco2021legio} for MPI applications, which deals with faults through graceful degradation, pointing out its limitations when dealing with the stencil pattern. We then circumvent those limitations, introducing the decoupled topology concept and the communication mechanisms for an incomplete topology. We then evaluate the additional functionalities on the iPiC3D application~\cite{Markidis2010}, highlighting the changes in the source code necessary to make the application fault-aware. We execute the application integrated with Legio in an experimental campaign on a production HPC cluster, showing the impact on the code result and execution times. Our work shows that faults affect the execution, but they only partially compromise the correctness of the results.

To summarise, the contributions of this paper are the following:
\begin{itemize}
\item We analyse the Legio fault resilience framework for MPI, highlighting its limitations in handling stencil applications;
\item We study the decoupled topology concept and define four strategies to deal with communication in incomplete topologies;
\item We evaluate iPiC3D as a specimen for stencil applications, and we evaluate the changes needed to make it fault-aware;
\item We explore the impact of faults on the fault-aware version of the iPiC3D application, showing the result's relevance despite the errors.
\end{itemize}

This paper is structured as follows: Section~\ref{sec:background} analyses the existing efforts in the literature dealing with faults in HPC. Section~\ref{sec:topo} then analyses the applicability of the Legio fault resilience framework for stencil applications. Section~\ref{sec:pic} describes the study on the iPiC3D application and the changes needed to make it fault-aware. Section~\ref{sec:experimental} illustrates the experimental campaign we conducted to measure the effect of faults on the iPiC3D application. Lastly, Section~\ref{sec:conclusion} concludes the paper.

\section{Background and Related Work}\label{sec:background}

This Section covers the main developments present in the literature in the context of fault management in MPI. Section~\ref{sec:background:ULFM} covers the ULFM extension and explains how a user can leverage its functionalities. Section~\ref{sec:background:stencil} discusses efforts that leveraged ULFM functionalities to introduce fault tolerance in MPI stencil applications. Section~\ref{sec:background:legio} overviews Legio, its strengths and differences from the other fault management frameworks.

\subsection{The ULFM Extension}\label{sec:background:ULFM}

The MPI standard does not include a method to recover the execution after detecting a fault. Its latest versions tried to minimise the impact faults cause, introducing Sessions~\cite{holmes2016mpi} for fault domain isolation and preventing error propagation from local calls (not requiring communication between processes). Nonetheless, an error originating from a communication call still compromises the entire MPI layer (or session if using the Session model), and the user cannot repair the damage.

The User Level Fault Mitigation (ULFM) extension~\cite{bland2013post} fixes this lack of functionality by providing the user with powerful tools to manage faults and their impact on executions. ULFM handles failures manifesting as abrupt termination of processes, providing functions for their detection, propagation and removal from execution. Using ULFM functionalities, the user can create fault-resilient MPI applications (i.e., able to continue past fault detection), and, with the help of state retrieval mechanisms like C/R, fault-tolerance (i.e., the effect of faults becomes null) is achievable too.

Fault detection mechanisms reside inside ULFM, and the user does not have to poll the system to check for their presence; on the contrary, MPI functions can return the new error code \texttt{MPI\_ERR\_PROC\_FAILED}, indicating the failure of some processes inside the communicator used. The user should always check the return code of the invoked functions and be ready to handle faults whenever the error code arises. 

The first step in the ULFM mechanism to deal with faults is their propagation: the rest of the repair procedure requires participation from all the processes part of the communicator exposing the errors (i.e., it is a collective operation). Processes mark communicators with faults using the function \texttt{MPIX\_Comm\_revoke}, to achieve fault propagation. Marked (\textit{revoked}) communicators become unusable for normal MPI operations, returning the error code \texttt{MPI\_ERR\_REVOKED} every time, but it is still possible to repair them. The revokedness of a communicator spreads among the processes part of it, and eventually, they will all be ready for the next step of the repair procedure.

After achieving a shared view on the necessity of repair, processes can call the \texttt{MPIX\_Comm\_shrink} function to remove failed ones from the communicator. The function operates analogously to a \texttt{MPI\_Comm\_split} and generates a working MPI communicator as if the fault never occurred. The new communicator may contain fewer processes than the original one (due to failures), but it preserves the rank order. If needed, the user can then introduce new processes inside the communicator by spawning them or leveraging spare ones, using standard MPI functionalities in both cases.

ULFM also provides functions to make processes agree on a shared return code (usually required for collective operations) and fetch the presence of known failures (fundamental for point-to-point operations with unspecified ranks). We will not discuss those features since we do not use them in our analysis.

\subsection{ULFM and Stencil Applications}\label{sec:background:stencil}

The ULFM extension functionalities are likely sufficient to deal with fault presence in any MPI application. Nonetheless, ULFM does not offer a standard method to recover the data and the status of the failed processes: the decision is deliberate since the best strategy depends on the application characteristics. The most explored direction is the one leveraging Checkpoint and Restart (C/R), with many efforts analysing the frequency, position and data to save periodically. Stencil applications benefit from these analyses since they fit the checkpoint recovery strategy well. This compatibility arises from stencil applications being iterative, and all the MPI data movements happen within an iteration, making its end the perfect moment to store a checkpoint. 

In the literature, we have multiple examples of efforts using C/R and ULFM for stencil applications: Teranishi et al.~\cite{teranishi2014toward} analysed MiniFE, a parallel finite element analysis code for thermal Partial Differential Equations (PDEs), which is also part of the Mantevo benchmark suite~\cite{heroux2009improving}. On the other hand, Losada et al.~\cite{losada2017assessing} considered the Himeno benchmark, a Poisson equation solver using the Jacobi iteration method. Both applications feature stencil characteristics and are feasible benchmarks to prove the applicability of ULFM and the C/R strategy to similar applications.

Gamell et al.~\cite{gamell2015local} focused on stencil applications, considering the possibility of using local rollback (i.e., restarting only failed processes). They experimented with a 1-D PDE solver and the S3D Stencil 3-D PDE solver, focusing on the impact of faults on the wall clock time. While starting from the ULFM extension functionality, their developed framework did not use ULFM due to the shrink operation's poor scalability.

Additionally, other efforts do not directly focus on stencil applications but whose results apply to the use case~\cite{pauli2015intrinsic}. While having some differences, they share a similar pattern for the removal of the fault damage: after fault detection, execution proceeds with the substitution of the damaged MPI data structure (either by ULFM shrink, re-initialisation through Reinit, or other methods), reconstruction of failed processes, and checkpoint load (either global or local), so to reach a consistent state. In the following subsection, we explore a framework focusing on fault resiliency, thus employing a different procedure.

\subsection{The Legio Fault Resilience Framework}\label{sec:background:legio}

The previous subsection discussed many efforts to introduce fault management properties through C/R. While C/R is compatible with all the applications, some of them may benefit from alternative approaches. In the literature, many efforts explored the possibility of leveraging application characteristics to reduce the amount of data to save/restore: in particular, data redundancy and Algorithm-Based Fault Tolerance (ABFT)~\cite{du2012algorithm, chen2008algorithm, duarte2014vcube} can simplify the recreation of a consistent state with benefits in terms of execution times and energy spent.

All these solutions allow the execution to return to a consistent state, nullifying the effect of the faults on the result. We refer to this achievement as fault tolerance. While prominent, it is not the only method to handle faults: it is possible to sacrifice result accuracy for a faster recovery, thus achieving fault resilience through graceful degradation. The Legio fault resilience framework~\cite{rocco2021legio} explores this direction with the idea of simplifying the creation of MPI applications that can deal with faults. The main idea behind the Legio framework is to continue the execution only with the processes which did not fail. This approach affects the accuracy of the result, but it may be an acceptable tradeoff for some applications, like Montecarlo solvers, where the work of each process is independent of the others. Moreover, due to not requiring the recreation of the failed process or the retrieval of its data lost with the fault, Legio reaches a faster recovery time than other fault tolerance solutions.

The design of Legio followed three core principles~\cite{rocco2023poster}: the developers want to preserve the scalability and performance of the applications using it (Efficiency) while limiting to the minimum their usage of MPI (Flexibility) and the number of code changes needed for supporting it (Transparency). The last principle is fundamental since HPC applications are usually large and stable, so changing them is non-trivial and cumbersome. By leveraging the PMPI layer, Legio integrates seamlessly with the application through linking, thus without making code changes in most cases. During the execution, Legio catches all the calls to MPI functions done by the application and substitutes the MPI data structures passed as parameters with copies. This substitution ensures that the MPI data structures handled by the application are not involved with faults, as they are never used inside the MPI functions. Legio repairs the copies it handles upon fault, not affecting the application versions.

Legio also provides a small API in case the application wants to query the status of its execution: it is fundamental to understand which processes failed and to plan recovery policies accordingly. The API also contains functions to deal with critical processes, i.e., those that must finish their execution to produce a meaningful result~\cite{rocco2024extending}. Using the Legio API is not mandatory, as users can leverage most of the functionality just by linking the framework to the application. However, the authors suggested the importance of fault awareness in MPI applications~\cite{rocco2023exploit}, stating that users can limit the loss of accuracy if the code is aware that some processes may stop responding. This assumption is not present in the current version of the MPI standard (v4.1), as it does not consider the behaviour after some process stops responding. Through Legio, the execution continues without changing ranks or communicators (all the changes happen within Legio), but we must consider that some processes may not perform some MPI calls, effectively changing the results. The authors of Legio already analysed a Montecarlo application~\cite{prahl1989monte}, showing the need for some minor changes to make it fault-aware and reduce the impact on accuracy. In this effort, we continue that analysis considering stencil applications, showing how a fault resilience approach can still produce meaningful results with minimal recovery times, even for tightly coupled applications.

\section{Fault Resilience for Stencil Applications}\label{sec:topo}

Due to its regularity and simplicity, the stencil communication pattern is very used in HPC applications. It usually involves a 2D or 3D space simulation; each process manages a small portion of the problem. At each iteration, each process computes the values of the simulated phenomenon in the portion of its competence using the data of its and the neighbour portions. Afterwards, it communicates the halo regions (i.e., the borders of its portion of competence) to its neighbours so they can use them for their following computations. In exchange, it receives the halo of the neighbours, which will be fundamental for the subsequent iteration.

Writing a stencil application in MPI is easy, as it mainly leverages data exchanges between pairs of neighbour processes. The MPI standard provides point-to-point operations that can fulfil data exchange needs. Moreover, it offers virtual process topologies, especially cartesian ones, which are a helpful abstraction that simplifies the development of regular applications. Using cartesian virtual topologies, users can map processes part of a communicator to a multidimensional grid or torus. The MPI standard provides methods to handle the mapping between the processes' ranks and coordinates in the virtual topology space, simplifying the handling of translations between the two. The participants of the US Exascale Computing project produced a survey on MPI usage in current and future HPC systems~\cite{bernholdt2020survey}: 11\% of the interviewed users leverage process topologies in their code, and they expected this number to grow to 21\% in the just-started exascale era.

To define a cartesian process topology, users must provide the number of dimensions, their size, and eventual periodicity to the MPI function \texttt{MPI\_Cart\_create}. The function generates a new communicator featuring cartesian virtual process topologies functionalities. Users can leverage those functionalities to simplify the localisation of processes in the space. A typical usage case is the movement of data along dimensions: the MPI standard defines the \texttt{MPI\_Cart\_shift} function, which receives the dimension to follow and the displacement as a parameter, and it computes the source and destination rank. The user can pass these ranks to a \texttt{MPI\_Sendrecv} call, shifting data along the chosen dimension following the correct displacement.

\subsection{Virtual Cartesian Topologies in Legio}

While cartesian virtual process topologies are very useful for stencil applications, their behaviour becomes problematic when faults occur. The \texttt{MPIX\_Comm\_shrink} operation, fundamental for the repair of damaged communicators, does not propagate the data regarding virtual process topologies. Moreover, the cartesian virtual process topology may be impossible to recreate after the repair procedure, as it requires a minimum number of processes (at least the product of the dimensions' sizes), and we may not have enough of them after the shrink operation. For this reason, if we want to pursue the fault resilience approach implemented by Legio, we must introduce some workaround.

The issue we face with virtual process topologies is rooted in the strict bond between them and communicators. In particular, it is only possible to define a virtual process topology within a communicator, and we lose this information when we free the communicator, even for repair purposes. If we break this dependency, we could store the virtual process topology information while changing the communicator, effectively propagating it through communicators. 

To support virtual process topologies in Legio, we follow the assumption above and decouple virtual process topologies from communicators, storing the data about the first in a separate data structure that preserves upon repair. Upon issuing virtual process topologies MPI operations, we use the information stored in the data structure, not the one in MPI communicators: we can achieve this since Legio operates at the PMPI level, allowing us to re-implement those functions. Upon repair, we substitute the communicator handled by Legio as usual, but we do not affect the virtual topology data structure, propagating it to the repaired communicator. This solution ensures the use of virtual process topologies even when we want fault resilience features, but we must still consider the effect failed processes may have on communication.

\subsection{Communication in an Incomplete Topology}

The Legio fault resilience framework has supported point-to-point communication since its first version~\cite{rocco2021legio}. Stencil applications mostly use point-to-point operations, and we must specify their behaviour in the presence of faults. In particular, we must consider two cases: a process sending data to a failed one, thus not affecting the computation, and a process expecting data from a failed one, therefore operating on wrong or invalid data. The second issue is the most critical of the two since it may lead to the propagation of the fault: a process using invalid or wrong data can access or modify areas of memory it should not do, thus incurring segmentation faults and stopping its execution. 

Legio deals with faulty point-to-point operations by not executing them if the process we are sending to/receiving from fails. This solution enables the continuation of the execution, and the process will act as if its neighbour shared what was originally in the receive buffer: if it is uninitialised, it may be random data leading to a potential fault. If we initialise the buffer with some valid value, we remove this eventuality: this is the core idea behind the first \textit{default} solution we propose. The default solution is simple to implement and presents minimal overhead, but it always provides the same value, introducing a drift towards it in its computations.

An alternative to the default solution is the \textit{mirror} approach: instead of relying on a constant value, we fill the buffer with the data sent to the failed process. Through the mirror approach, we get the same benefits as the default alternative but with varying values, thus removing the drift towards the default. On the other hand, the reflection effect coming from the mirror affects the correctness of the result, and it may create singularity points on the corners of the failed process (as neighbours coming from different directions see only their data, not the one of the other).

\begin{figure}[t]
    \centering
    \includegraphics[width=\linewidth]{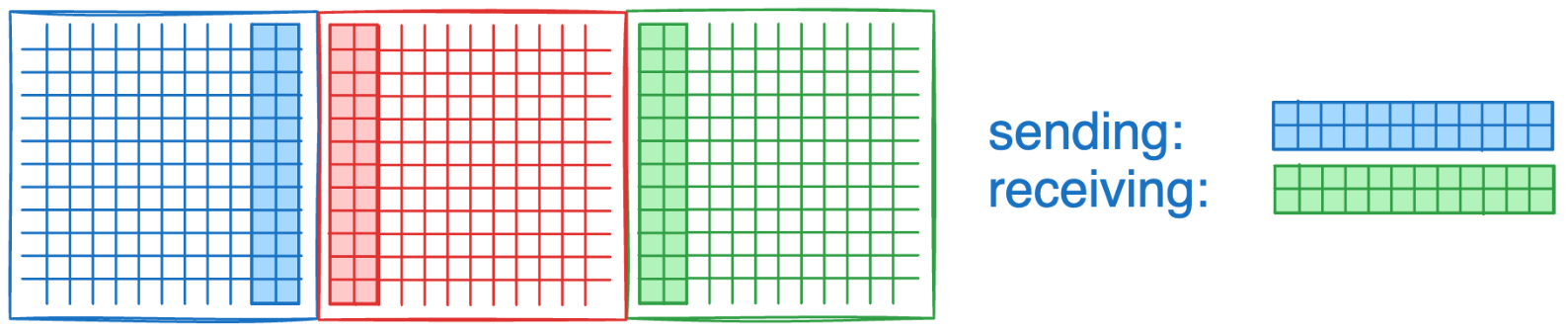}
    \caption{Bridge solution to deal with the absence of a process (in red) in the communication seen from the perspective of the blue process. The process receives the data sent by the neighbour's neighbour.}
    \label{fig:bridge}
\end{figure}

While the default and mirror cases did not use data coming from other processes, with the \textit{bridge} solution, we want to explore that possibility. In particular, a process expecting data from a failed one can receive it from the one that would have sent it to the failed one, creating a sort of bridge (hence the name) over the failure. Figure~\ref{fig:bridge} represents this idea, where data overcomes a fault in the topology. By bridging, we do not discard the value sent to a failed process, but we reuse it, sending it to the next one. On the other hand, bridging changes the topology, damaging some geometrical properties (data travels faster along the dimension featuring a fault), and it may be worse for result accuracy.

Implementing the bridging strategy is not straightforward, as we must select a new neighbour with whom to communicate. In Legio, we implemented the bridging strategy by redefining the behaviour of the \texttt{MPI\_Cart\_shift}, which will point to the following non-failed process along the dimension. With this change, the function can produce different results at different times, so it becomes mandatory to use it every time before calling the \texttt{MPI\_Sendrecv} operation. 

Additionally, users can combine the bridging and default approach to estimate the value handled by the failed process through \textit{interpolation} of the values at both ends of the fault. This approach is trickier since it requires deep knowledge of the data exchanged between processes, but it can provide significant benefits regarding result accuracy. We will not explore this direction since it is too application-specific and intrusive inside the code. All the other solutions are instead available within the Legio framework, and the user can select the one that fits the application code the most and reduces the damage to the application results.



\section{The iPiC3D application use case}\label{sec:pic}

The plasma physics code iPiC3D~\cite{Markidis2010} is a paramount example of a stencil communication pattern. iPiC3D is a Particle-in-Cell (PiC) or particle-mesh code~\cite{Dawson1983}. The PiC algorithm adopts a statistical approach and models plasmas as ensembles of computational particles or macroparticles, which can be seen as small pieces of phase space~\cite{Lapenta2011} or blobs of incompressible phase fluid moving in phase space~\cite{Gibbon2007}. Plasma macroparticles interact self-consistently through the electromagnetic fields that they produce. These fields are computed on a fixed grid by solving Maxwell's equation, where source terms are obtained by depositing discrete particles into the grid. PiC codes are generally used to explore the plasma dynamics at a kinetic level. In particular, iPiC3D is widely adopted to model space plasmas. These plasmas are predominately collisionless, and their components (electrons, protons, and heavier ions) show distribution functions often far from Maxwellian equilibrium. Hence, they can only be correctly modelled by adopting a kinetic approach. Among the different possibilities, the PiC method is likely the most practical due to its limited number of physics approximations and the limited (compared to other solutions) amount of computational resources necessary to run. In addition, iPiC3D implements the Implicit Moment Method~\cite{Brackbill1982}. This scheme relaxes some of the severe constraints on the spatial and temporal resolution characteristics of traditional PiC algorithms, thus making iPiC3D particularly suitable for investigating multiscale plasmas processes typical of the heliosphere.

The code iPiC3D is written in C/C++ and parallelised with MPI. The total physical domain is divided among the processes following a Cartesian topology. Each process controls the field values in its subdomain and the particles within. 

Discretised Maxwell's equations form a non-symmetric linear system, which is solved using the Generalised Minimal Residual (GMRes) method~\cite{Saad1986}. The library PETSc~\cite{petsc-web-page} can be utilised for better performance. However, a homegrown version of the solver is also present.

iPiC3D saves data using the parallel hdf5 library~\cite{koranne2011}. The code uses CMake as a building system and is available on GitHub\footnote{\url{https://github.com/CmPA/iPic3D}}.

\subsection{Introducing Legio in iPiC3D}

While Legio does not require code changes for its introduction, it still needs the user to assess the damage that faults may cause in the execution and plan some countermeasures correctly~\cite{rocco2024overview}. Limiting ourselves to linking the Legio library against the application may not produce the desired results as the algorithms may leverage some valid assumptions in a fault-free scenario that may become wrong. We performed this analysis on the iPiC3D application to assess the usage of Legio for stencil applications.

Being a PiC application, iPiC3D follows two different communication patterns: on one side, we have the classical stencil pattern for the computation of the electric and magnetic fields; on the other, we have the simulation of particles in space. The communication follows cartesian topologies, exchanging data with neighbour processes. For field computation, processes exchange halos (the border region of the spatial area computed by each process) containing the values of the fields, while for particle movement, processes exchange particles moving from one region to another. 


The communication between pairs of neighbour processes happens using the MPI function \texttt{MPI\_Sendrecv\_replace}, which operates analogously to the \texttt{MPI\_Sendrecv} one but requires only a single buffer. The buffer must contain the data to send upon calling the function and will hold the received data upon returning, thus overwriting the sent values. In the application code, the source and destination ranks specified in the functions are always identical: this implies that processes exchange information with a neighbour at a time rather than shifting data along dimensions. This structure also means that if we do not perform the call (due to the neighbour's fault), the process will continue working on the data it was supposed to send, innately achieving the \textit{mirror} approach. We decided to rely on this innate data management method as it is the one closest to the natural application behaviour and is available without code changes.

\subsection{Fault-awareness for the iPiC3D application}

Continuing our code analysis, we see that the particle-moving part of the application requires minor refactoring. The original algorithm for particles moving across the topology analyses the coordinates of each particle and compares them to the range of values handled by the process. If one dimension is smaller than the minimum handled or greater than the maximum, the process sends the particle to the previous or following process in that direction, respectively. The algorithm performs this check on the three spacial dimensions in order, and a particle may need multiple movements to reach its intended spot in the topology. This strategy is valid as the receiving process will repeat these checks so that a particle moving diagonally in space can hop two processes (first along the x coordinate and then along the y coordinate) and reach its destination.

\begin{figure}[t]
    \centering
    \includegraphics[width=0.6\linewidth]{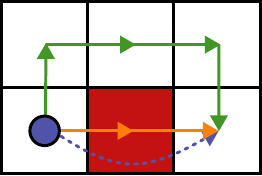}
    \caption{Comparison between particle movements in the presence of faults with the original and A* algorithms. The particle (the blue circle) should move from the bottom-left process to the bottom-right (following the blue dashed arrow), with a fault on the center-bottom (the one in red). The original algorithm tries to move the particle on the x-axis (following orange arrows), but it cannot due to the fault. The A* algorithm (green arrows) overcomes the limitation by sending the particle upwards first.}
    \label{fig:diagonal}
\end{figure}

In the presence of faults, that algorithm becomes insufficient: process faults imply that the forwarding may not always work, as particles cannot traverse a failed process. There may exist a path to the process that will take care of the particle, but considering only the coordinates without a holistic view of the topology does not allow it. Figure~\ref{fig:diagonal} shows the issue faced by the algorithm: the particle must move along the x direction by two processes, but the first hop leads to a failed process; thus, it cannot reach the destination using the original algorithm. However, if we look at the entire topology, we see a path to the destination that avoids the fault: we could follow that, but the original algorithm cannot find it. In general, switching the algorithm to an A*~\cite{hart1968formal} implementation solves the issue, as it allows particles to reach the destination as long as a path exists, even with some faults. 

While A* allows us to circumvent fault presence, its implementation is more cumbersome than the original algorithm, both in terms of code lines and execution time: for this reason, we execute it only if faults may interfere with the path to the destination of the particle. We use the A* algorithm only to select the direction to forward the particle, ensuring that there exists a path from there to the destination, but we do not pass the path alongside the particle, as it would alter the communication pattern of the application significantly. Alongside adopting A*, the application must recognise the loss of a process and discard all the particles heading toward it. This solution is mandatory since processes always try to forward particles not belonging to their space, but the ones belonging to failed processes do not have a correct recipient and thus would move around forever.

After employing these changes, the application correctly interacts with Legio and can leverage all the fault resilience functionalities that come from it. Nonetheless, we still need to address the damage on result accuracy from the absence of one (or more) process. The following section assesses the potential of Legio for fault handling in the iPiC3D application through two real-world use case scenarios.

\section{Experimental campaign}\label{sec:experimental}

The following experimental campaign evaluates the impact on the results of the iPiC3D application caused by faults when employing fault resilience. In particular, we want to show that some faults may not affect the results produced. Thus, we can avoid rolling back to a previously saved checkpoint in those cases. It is mandatory to point out that we do not address a generic fault, as its location in the topology significantly affects the accuracy loss. Nonetheless, by allowing the application to continue past fault presence, we also enable a tradeoff between result correctness and time to solution, as restarting the execution is way more costly than continuing with fewer processes. 

We execute our experiments on the EuroHPC Karolina cluster managed by IT4Innovations, featuring nodes with 2 x AMD Zen 2 EPYC™ 7H12, 2.6 GHz processors, and 256 GB of RAM. Each node can run up to 128 processes without overloading, but we limit this number to 64 to better spread the RAM available between the processes. We compile our code using OpenMPI v5.0.0 featuring ULFM, implementing MPI standard v4.1. In faulty executions, we inject faults by making selected processes raise a \texttt{SIG\_INT} signal.

The two experiments we consider represent actual use cases of the iPiC3D simulation. The first scenario analyses the Weibel instability phenomenon~\cite{weibel1959spontaneously}, while the second recreates the magnetic reconnection process~\cite{giovanelli1946theory, giovanelli1947magnetic, mozer2010magnetic}. We analyse execution times and result accuracy, with the latter based on a metric of interest for each experiment. We inject faults into single processes and analyse their impact on the metrics of interest. We must remark that fault position \textbf{does} highly affect the goodness of the results: this experimental campaign aims not to prove that the application is resilient to any fault but rather to highlight that introducing fault resilience can produce benefits in some cases. 

\subsection{The Weibel instability experiment}

\begin{figure*}[t]
    \centering
    \begin{subfigure}[]{0.47\linewidth}
        \centering
        \includegraphics[width=\linewidth]{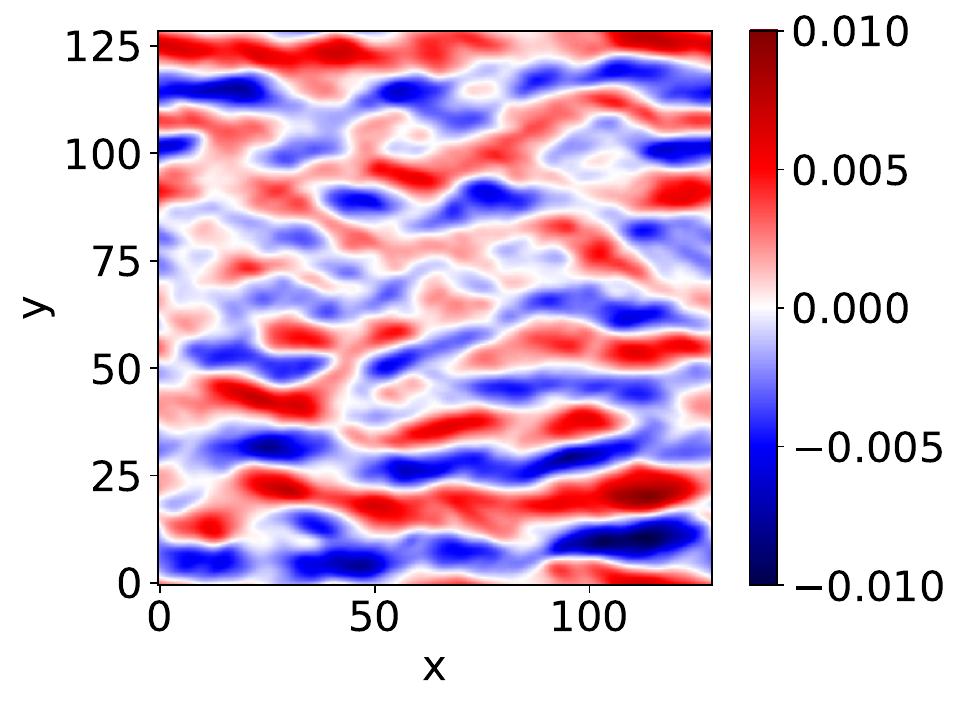}
        \caption{Fault free}
        \label{fig:weibel_ok}
    \end{subfigure}
    \hfill
    \begin{subfigure}[]{0.47\linewidth}
        \centering
        \includegraphics[width=\linewidth]{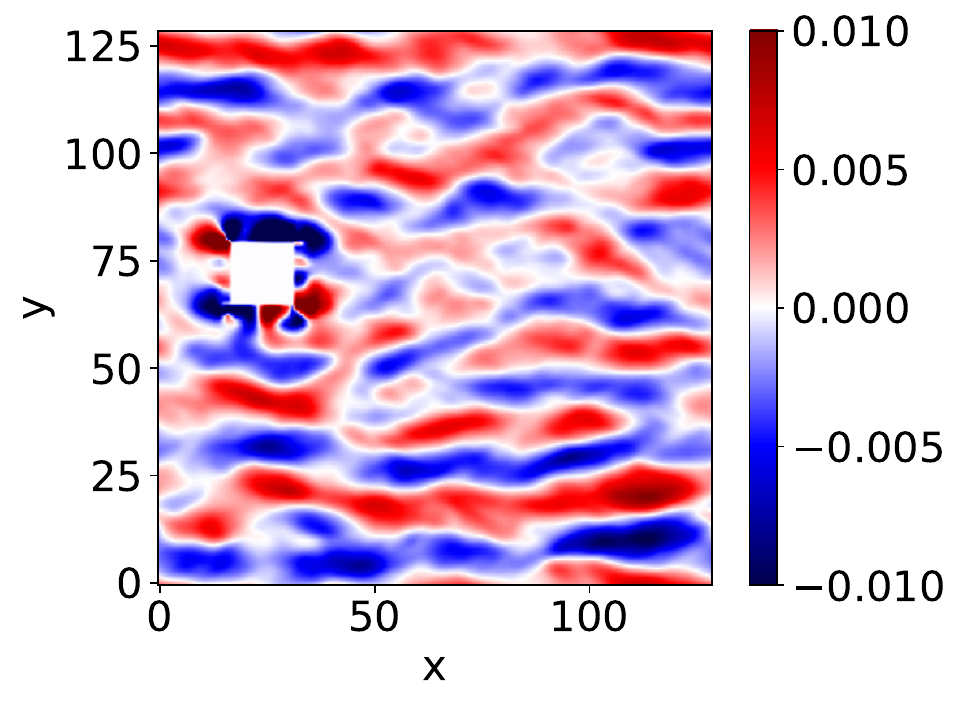}
        \caption{Faulty}
        \label{fig:weibel_ko}
    \end{subfigure}
    \caption{Comparison between the executions of the Weibel instability experiment in the absence and presence of faults respectively. These plots show the intensity of the Magnetic field along the z-axis at iteration 320 (40 after the fault). The unit of measure of the field is $\frac{m_{i} c \omega_p}{e}$, where $m_i$ is the mass of a proton, $e$ is the elementary charge, $c$ is the speed of light in vacuum, and $\omega_{p}$ is the plasma frequency of the protons. Axis are normalised to a factor of $c/\omega_p$.}
    \label{fig:weibel}
\end{figure*}

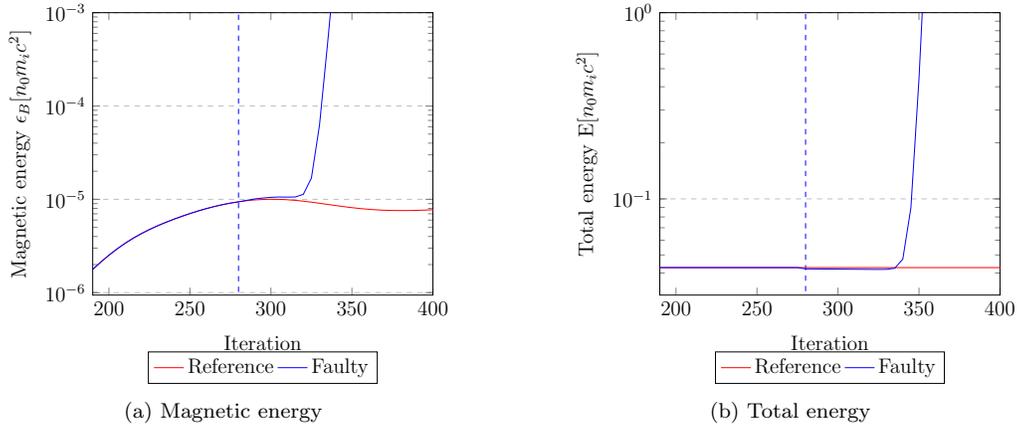
\begin{figure*}[t]
    \centering
    \begin{subfigure}[]{0.45\textwidth}
        \centering
        \pgfplotsset{
/pgfplots/boxplot legend/.style={
   legend image code/.code={
     \draw [|-|,##1] (0,1.5mm) --
         node[rectangle,minimum size=2mm,draw,fill,fill opacity=1,##1]{}
        (0,5mm);
    }
  }
}


\begin{tikzpicture}[scale=0.6,transform shape]

\begin{axis}[
    title style={at={(0.5,-0.3)},anchor=north},
    xlabel style={at={(0.5,-0.12)},anchor=north},
    xlabel={Iteration},
    ylabel={Magnetic energy $\epsilon_B$[$n_0 m_i c^2$]},
    ymode=log,
    xmin=190,
    xmax=400,
    ymax=0.001,
    legend style={at={(0.5,-0.2)},
    anchor=north,legend columns=-1},
    ymajorgrids=true,
    y grid style=dashed,
]

\addlegendentry{Reference}
\addlegendentry{Faulty}
\draw[dashed,blue] (280,0.0000001) -- (280,0.01);

\addplot[color=red] plot coordinates {
(0,2.73487654556794e-10)
(5,3.51853453504009e-09)
(10,1.99572800489207e-09)
(15,1.67105847852141e-09)
(20,2.17161863552587e-09)
(25,2.71321621651958e-09)
(30,3.14650490557445e-09)
(35,3.66673373527099e-09)
(40,4.34274786656143e-09)
(45,5.10860241505827e-09)
(50,5.95798033764648e-09)
(55,6.95013919496727e-09)
(60,8.18321809484571e-09)
(65,9.76446531313669e-09)
(70,1.17445525742616e-08)
(75,1.42242775888359e-08)
(80,1.7395846162352e-08)
(85,2.1473389538537e-08)
(90,2.67186397585479e-08)
(95,3.34125702442712e-08)
(100,4.19774873940199e-08)
(105,5.30377550043813e-08)
(110,6.6966031646286e-08)
(115,8.43186519737037e-08)
(120,1.05680802926229e-07)
(125,1.31750577530229e-07)
(130,1.63251239359863e-07)
(135,2.00855671244921e-07)
(140,2.45212001399852e-07)
(145,2.97619825529559e-07)
(150,3.60270717727594e-07)
(155,4.36694527243867e-07)
(160,5.30629390317089e-07)
(165,6.47385368033457e-07)
(170,7.93329915496436e-07)
(175,9.74958927476133e-07)
(180,1.19787054585121e-06)
(185,1.46454412105268e-06)
(190,1.77524125799084e-06)
(195,2.12833028254808e-06)
(200,2.51905678476936e-06)
(205,2.94024826102214e-06)
(210,3.38147885246201e-06)
(215,3.83365629720073e-06)
(220,4.29036794270314e-06)
(225,4.74730089739256e-06)
(230,5.2023483356692e-06)
(235,5.65732912661035e-06)
(240,6.11499059988983e-06)
(245,6.57645332921366e-06)
(250,7.040306634317e-06)
(255,7.50044930433969e-06)
(260,7.94652139745706e-06)
(265,8.3691443205988e-06)
(270,8.76105844027161e-06)
(275,9.11516837584319e-06)
(280,9.42138181558572e-06)
(285,9.67023469620333e-06)
(290,9.8555227514711e-06)
(295,9.97317773676151e-06)
(300,1.00200200838269e-05)
(305,9.99581164800972e-06)
(310,9.90439207740018e-06)
(315,9.75407030107781e-06)
(320,9.55672816602881e-06)
(325,9.32603810865927e-06)
(330,9.07774088746536e-06)
(335,8.82518213188243e-06)
(340,8.57975200265989e-06)
(345,8.35144761403697e-06)
(350,8.14814545588263e-06)
(355,7.97450076967566e-06)
(360,7.83124362739402e-06)
(365,7.72021904891685e-06)
(370,7.64320637198773e-06)
(375,7.59828745154091e-06)
(380,7.58206340976052e-06)
(385,7.59106078164621e-06)
(390,7.62187817434055e-06)
(395,7.67039212503072e-06)
(400,7.73337570081301e-06)
(405,7.80880423834258e-06)
(410,7.89623817672338e-06)
(415,7.99448489194507e-06)
(420,8.10291323343039e-06)
(425,8.22075124403652e-06)
(430,8.34651626309412e-06)
(435,8.47974318741317e-06)
(440,8.61958610071044e-06)
(445,8.76298869592757e-06)
(450,8.90638651516676e-06)
(455,9.04695703143893e-06)
(460,9.18262202877179e-06)
(465,9.31284738941807e-06)
(470,9.4379622950333e-06)
(475,9.55932839896416e-06)
(480,9.67751465342389e-06)
(485,9.79297691244244e-06)
(490,9.90441812771678e-06)
(495,1.00121217008165e-05)
(500,1.01175112804475e-05)
(505,1.0222720524033e-05)
(510,1.03314057825614e-05)
(515,1.04470586344996e-05)
(520,1.05730536062042e-05)
(525,1.07107429620508e-05)
(530,1.08598581436252e-05)
(535,1.10185750664654e-05)
(540,1.11848288713091e-05)
(545,1.13567858118851e-05)
(550,1.15325629906262e-05)
(555,1.17118551401272e-05)
(560,1.18951770413784e-05)
(565,1.20830593956797e-05)
(570,1.22760004819181e-05)
(575,1.24740907772429e-05)
(580,1.26783827755375e-05)
(585,1.28896762461949e-05)
(590,1.31079522227291e-05)
(595,1.33339005210224e-05)};

\addplot[color=blue] plot coordinates {
(0,2.73487654556794e-10)
(5,3.51853453504009e-09)
(10,1.99572800489207e-09)
(15,1.67105847852141e-09)
(20,2.17161863552587e-09)
(25,2.71321621651958e-09)
(30,3.14650490557445e-09)
(35,3.66673373527099e-09)
(40,4.34274786656143e-09)
(45,5.10860241505827e-09)
(50,5.95798033764648e-09)
(55,6.95013919496727e-09)
(60,8.18321809484571e-09)
(65,9.76446531313669e-09)
(70,1.17445525742616e-08)
(75,1.42242775888359e-08)
(80,1.7395846162352e-08)
(85,2.1473389538537e-08)
(90,2.67186397585479e-08)
(95,3.34125702442712e-08)
(100,4.19774873940199e-08)
(105,5.30377550043813e-08)
(110,6.6966031646286e-08)
(115,8.43186519737037e-08)
(120,1.05680802926229e-07)
(125,1.31750577530229e-07)
(130,1.63251239359863e-07)
(135,2.00855671244921e-07)
(140,2.45212001399852e-07)
(145,2.97619825529559e-07)
(150,3.60270717727594e-07)
(155,4.36694527243867e-07)
(160,5.3062939031709e-07)
(165,6.47385368033456e-07)
(170,7.93329915496436e-07)
(175,9.74958927476133e-07)
(180,1.19787054585121e-06)
(185,1.46454412105268e-06)
(190,1.77524125799085e-06)
(195,2.12833028254808e-06)
(200,2.51905678476936e-06)
(205,2.94024826102214e-06)
(210,3.38147885246201e-06)
(215,3.83365629720074e-06)
(220,4.29036794270314e-06)
(225,4.74730089739257e-06)
(230,5.20234833566921e-06)
(235,5.65732912661036e-06)
(240,6.11499059988984e-06)
(245,6.57645332921367e-06)
(250,7.04030663431701e-06)
(255,7.5004493043397e-06)
(260,7.94652139745707e-06)
(265,8.36914432059881e-06)
(270,8.76105844027162e-06)
(275,9.1151683758432e-06)
(280,9.3862898363513e-06)
(285,9.75411626245247e-06)
(290,1.00978812612889e-05)
(295,1.03271128401365e-05)
(300,1.05079550964651e-05)
(305,1.05960901779633e-05)
(310,1.05711439757838e-05)
(315,1.05951101612978e-05)
(320,1.1312408794075e-05)
(325,1.6917167799816e-05)
(330,6.26008913361204e-05)
(335,0.000472858751798361)
(340,0.00399529850080336)
(345,0.0333294554485672)
(350,0.276052456454579)
(355,2.28944765633632)
(360,18.9934135057948)
(365,157.498984511258)
(370,1306.25923494591)
(375,10836.0476797821)
(380,89899.5988257641)
(385,745873.034923521)
(390,6188420.58828119)
(395,51344913.8031859)
(400,426006391.138614)};

    


\end{axis}

\end{tikzpicture}
        \caption{Magnetic energy}
        \label{fig:weibel_B}
    \end{subfigure}
    \hfill
    \begin{subfigure}[]{0.45\textwidth}
        \centering
        \pgfplotsset{
/pgfplots/boxplot legend/.style={
   legend image code/.code={
     \draw [|-|,##1] (0,1.5mm) --
         node[rectangle,minimum size=2mm,draw,fill,fill opacity=1,##1]{}
        (0,5mm);
    }
  }
}

\begin{tikzpicture}[scale=0.6,transform shape]

\begin{axis}[
    title style={at={(0.5,-0.3)},anchor=north},
    xlabel style={at={(0.5,-0.12)},anchor=north},
    xlabel={Iteration},
    ylabel={Total energy E[$n_0 m_i c^2$]},
    ymode=log,
    xmin=190,
    xmax=400,
    ymax=1,
    legend style={at={(0.5,-0.2)},
    anchor=north,legend columns=-1},
    ymajorgrids=true,
    y grid style=dashed,
]

\addlegendentry{Reference}
\addlegendentry{Faulty}
\draw[dashed,blue] (280,0.0000001) -- (280,1);

\addplot[color=red] plot coordinates {
(0,0.0427645601311896)
(5,0.04276454635002)
(10,0.0427645259536001)
(15,0.0427645044324145)
(20,0.0427644820123075)
(25,0.0427644584239408)
(30,0.0427644341375798)
(35,0.0427644086857248)
(40,0.0427643831356168)
(45,0.0427643570530361)
(50,0.0427643295171566)
(55,0.0427643010783039)
(60,0.0427642716483784)
(65,0.0427642408137925)
(70,0.0427642088920985)
(75,0.0427641752269081)
(80,0.0427641404144832)
(85,0.0427641032740383)
(90,0.0427640634048343)
(95,0.0427640184366088)
(100,0.0427639673358645)
(105,0.0427639086855936)
(110,0.0427638397639704)
(115,0.0427637591982389)
(120,0.0427636650372313)
(125,0.04276355741442)
(130,0.0427634390860255)
(135,0.0427633120244431)
(140,0.0427631836511726)
(145,0.0427630580063277)
(150,0.0427629389777483)
(155,0.0427628263897278)
(160,0.0427627227182557)
(165,0.0427626236892808)
(170,0.0427625267365062)
(175,0.0427624294104192)
(180,0.0427623297686958)
(185,0.0427622307394489)
(190,0.0427621318627861)
(195,0.0427620325973235)
(200,0.0427619368223542)
(205,0.0427618435863068)
(210,0.0427617583321638)
(215,0.0427616830832989)
(220,0.0427616168890924)
(225,0.0427615574051074)
(230,0.0427615031156366)
(235,0.0427614500641011)
(240,0.0427613994947531)
(245,0.0427613498866791)
(250,0.0427612998710127)
(255,0.0427612489920772)
(260,0.0427611990200229)
(265,0.0427611502877806)
(270,0.0427611023963558)
(275,0.0427610538063657)
(280,0.0427610052326669)
(285,0.0427609569335415)
(290,0.0427609097530072)
(295,0.0427608632917916)
(300,0.0427608189941754)
(305,0.042760777030952)
(310,0.0427607365977256)
(315,0.0427606985386676)
(320,0.0427606608628829)
(325,0.042760623731485)
(330,0.0427605866752026)
(335,0.0427605492277704)
(340,0.0427605115729172)
(345,0.042760473572239)
(350,0.0427604347753098)
(355,0.0427603945445368)
(360,0.0427603534148375)
(365,0.0427603108415237)
(370,0.0427602676816232)
(375,0.042760224020871)
(380,0.0427601787625766)
(385,0.0427601331623152)
(390,0.0427600876891502)
(395,0.0427600421730367)
(400,0.0427599948690115)
(405,0.0427599479797278)
(410,0.0427599008136883)
(415,0.0427598536759056)
(420,0.0427598072461449)
(425,0.0427597607753202)
(430,0.0427597142974173)
(435,0.0427596673630562)
(440,0.0427596204133308)
(445,0.0427595743351805)
(450,0.0427595297077269)
(455,0.0427594837262796)
(460,0.0427594419613608)
(465,0.0427593971107207)
(470,0.0427593529619687)
(475,0.0427593083190873)
(480,0.0427592632484096)
(485,0.0427592178161258)
(490,0.042759172056415)
(495,0.0427591272126493)
(500,0.0427590832834011)
(505,0.0427590390503202)
(510,0.0427589942560802)
(515,0.0427589493929925)
(520,0.0427589035668114)
(525,0.0427588580793308)
(530,0.042758813079522)
(535,0.0427587683575686)
(540,0.0427587232937178)
(545,0.0427586784275337)
(550,0.042758634011635)
(555,0.042758589371954)
(560,0.0427585448917386)
(565,0.0427585001631786)
(570,0.0427584561136579)
(575,0.0427584119460214)
(580,0.0427583675035616)
(585,0.0427583230366762)
(590,0.0427582779490901)
(595,0.0427582325455415)};

\addplot[color=blue] plot coordinates {
(0,0.0427645601311896)
(5,0.04276454635002)
(10,0.0427645259536001)
(15,0.0427645044324145)
(20,0.0427644820123075)
(25,0.0427644584239408)
(30,0.0427644341375798)
(35,0.0427644086857248)
(40,0.0427643831356168)
(45,0.0427643570530361)
(50,0.0427643295171566)
(55,0.0427643010783039)
(60,0.0427642716483784)
(65,0.0427642408137925)
(70,0.0427642088920985)
(75,0.0427641752269081)
(80,0.0427641404144832)
(85,0.0427641032740383)
(90,0.0427640634048343)
(95,0.0427640184366088)
(100,0.0427639673358645)
(105,0.0427639086855936)
(110,0.0427638397639704)
(115,0.0427637591982389)
(120,0.0427636650372313)
(125,0.04276355741442)
(130,0.0427634390860255)
(135,0.0427633120244431)
(140,0.0427631836511726)
(145,0.0427630580063277)
(150,0.0427629389777483)
(155,0.0427628263897278)
(160,0.0427627227182557)
(165,0.0427626236892808)
(170,0.0427625267365062)
(175,0.0427624294104192)
(180,0.0427623297686958)
(185,0.0427622307394489)
(190,0.0427621318627861)
(195,0.0427620325973236)
(200,0.0427619368223542)
(205,0.0427618435863068)
(210,0.0427617583321638)
(215,0.0427616830832989)
(220,0.0427616168890924)
(225,0.0427615574051074)
(230,0.0427615031156366)
(235,0.0427614500641011)
(240,0.0427613994947531)
(245,0.0427613498866791)
(250,0.0427612998710127)
(255,0.0427612489920772)
(260,0.0427611990200229)
(265,0.0427611502877805)
(270,0.0427611023963558)
(275,0.0427610538063657)
(280,0.0420866320513062)
(285,0.0420556996165673)
(290,0.0420250416564979)
(295,0.0419940756037672)
(300,0.0419626304592774)
(305,0.0419314552813347)
(310,0.0419003392400831)
(315,0.0418690910710726)
(320,0.0418386076922413)
(325,0.0418156045449285)
(330,0.0418526990249254)
(335,0.0424172439371896)
(340,0.047437332785915)
(345,0.0895293955013612)
(350,0.440699676722406)
(355,3.36964974133447)
(360,27.7188382069628)
(365,229.824203776412)
(370,1906.98398200712)
(375,15823.3430350089)
(380,131288.866663344)
(385,1089301.55952229)
(390,9037833.49282616)
(395,74985771.3064311)
(400,622146812.463939)};

    


\end{axis}

\end{tikzpicture}
        \caption{Total energy}
        \label{fig:weibel_energy}
    \end{subfigure}
    \caption{Energy measurements of the Weibel instability experiment. For reference, $n_0$ is the plasma density, $m_i$ is the mass of a proton, $c$ is the speed of light in vacuum, and $\omega_{p}$ is the plasma frequency of the protons.}
    \label{fig:w_energy}
\end{figure*}

For the Weibel instability scenario, we execute our application on a single node featuring 64 processes in an 8x8 grid. Each process handles a grid of 16x16 cells; thus, the simulation involves over 16 thousand cells. We let the application run for 600 iterations, and we are interested in measuring the variation of magnetic energy of the system, with a particular focus on the plateau it reaches after some iterations. For the faulty execution, we inject a fault into the process with rank 12 after 280 iterations from the beginning. Figure~\ref{fig:weibel_ok} and~\ref{fig:weibel_ko} compare the value of the z component of the magnetic field, and Figure~\ref{fig:weibel_B} plots the total magnetic energy density for both cases. While the fault has a remarkable impact on the execution close to the fault location, the magnetic energy plot varies after reaching the plateau of interest; thus, the execution still produces usable data. Soon after getting to the plateau, the faulty execution diverges exponentially, producing wrong values that are easy to detect: we can spot this by checking the energy conservation law and stopping the execution once the variation of the total energy overcomes a certain threshold. Figure~\ref{fig:weibel_energy} plots total energy for the execution with the fault: the value stays close to the fault-free reference until fault insurgence, then we see a sudden drop by about 1.5\%, and the values stabilise again. After a couple of iterations, the values start growing until they reach unbearable numbers close to iteration 355, way past the saturation of the instability occurring when the maximum value of the magnetic energy is reached. The initial drop is due to the loss of the energy of the failed process (indeed, we lost 1/64 of the execution, which evaluates roughly to 1.5\%), while the latter growth comes from the corruption of the values across the execution.

From all these data, we can see that the execution could withstand the fault presence for several iterations before producing meaningless data: it lasted for about 70 iterations, more than 10\% of the total execution. In particular, despite the fault of a process, data generated in the simulation are still meaningful to explore the linear phase of the instability (when the magnetic energy increases) and the saturation process (when the magnetic energy reaches the plateau), which are still open questions in plasma physics. The application generated the data we were interested in, thus turning a fault-compromised execution into a successful one, saving the time needed to reload the previous checkpoint and reschedule execution.

\begin{figure}[t]
    \centering
    \pgfplotsset{
/pgfplots/boxplot legend/.style={
   legend image code/.code={
     \draw [|-|,##1] (0,1.5mm) --
         node[rectangle,minimum size=2mm,draw,fill,fill opacity=1,##1]{}
        (0,5mm);
    }
  }
}

\begin{tikzpicture}[scale=0.6,transform shape]

\begin{axis}[
    title style={at={(0.5,-0.3)},anchor=north},
    xlabel style={at={(0.5,-0.12)},anchor=north},
    xlabel={Iteration},
    ylabel={Time [s]},
    xmin=0,
    xmax=370,
    legend style={at={(0.5,-0.2)},
    anchor=north,legend columns=-1},
    ymajorgrids=true,
    y grid style=dashed,
]

\draw[dashed,blue] (280,0) -- (280,200);

\addplot[color=blue] plot coordinates {};

\addplot [color=blue] table[y=diff,x=time,col sep=comma] {plots/small_timings.csv};

    


\end{axis}

\end{tikzpicture}
    \caption{Time to complete 10 iteration of the algorithm as a function of the iteration for the execution with faults of the Weibel instability experiment.}
    \label{fig:weibel_time}
\end{figure}
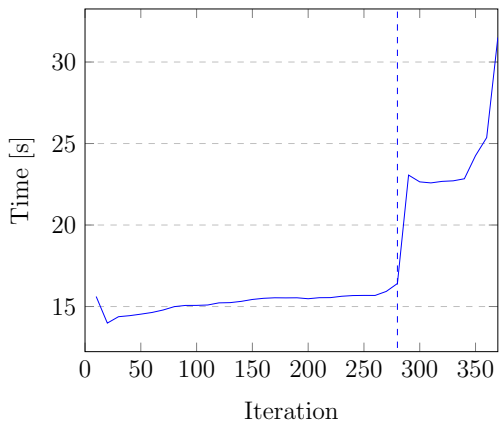

While this result strongly affirms the potential of integrating Legio within the iPiC3D application, we must also underline some limitations. First and foremost, the instant featuring the fault affects the usability of the result: given that the application with Legio is resilient for about 70 iterations past a fault, if the fault happens too early, we cannot analyse the plateau and must rerun the execution. Additionally, Legio changes the particle routing algorithm after fault insurgence, and it is possible to view this impact in Figure~\ref{fig:weibel_time}, where we compare the execution times of faulty and fault-free scenarios. After a fault, the time needed to move particles increases as we must change the algorithm to the more demanding A*. Additionally, the absence of the process dramatically perturbates the fields, thus accelerating particles in an anomalous way: to deal with particles moving this fast, we need many communication iterations, slowing down the process even more as we continue the execution past fault presence. Lastly, the absence of a process creates an anomaly in the fields, and the computation of the evolution of the fields takes more time due to the irregularity of the space. At a certain point, the execution just requires too much time to perform an iteration, so it is pointless to continue; nonetheless, this happens way after breaking the energy conservation law, so we should stop the execution before.

With this first test, we demonstrated the potential of Legio when integrated with the iPiC3D application. We executed this test on a relatively small MPI network, counting only 64 processes: while this may seem a limitation of the current experiment, it means that a single process fault causes more damage to the problem, as the percentage of area lost with the fault is more significant with fewer processes. Nonetheless, the following experiment involves a larger MPI topology to assess the impact of faults on scalability, too.

\subsection{The magnetic reconnection experiment}


With the second experiment, we simulate plasma behaviour in conditions that show the magnetic reconnection phenomenon. We execute this experiment using 16 Karolina nodes running 64 processes, obtaining a 1024-process network. Each process simulates 1024 cells. Thus, the entire simulation involves more than 1 million cells. The simulation runs for 3000 iterations. We want to measure the reconnection rate, i.e., the rate at which the magnetic flux undergoing the reconnection process changes. From a mathematical point of view, this quantity can be computed as the ratio between the y component of the electric field evaluated at the so-called X point, the point where magnetic field lines reconnect (in our case the centre of the top-right quadrant) and the product between the x component of the magnetic field and the Alfvén speed upstream of the X point. For this experiment, we are interested in measuring the spikes in the magnetic reconnection values. For the faulty execution, we inject a fault at iteration 1800 on process 67: in this case, not only does the time instant influence the results, but also the fault location (the closer the fault to the perturbation, the higher the impact on the result). Process 67 is far from the measured perturbation area, so its impact will affect the outcome after many iterations.

\begin{figure*}[t]
    \centering
    \begin{subfigure}[]{0.47\linewidth}
        \centering
        \includegraphics[width=\linewidth]{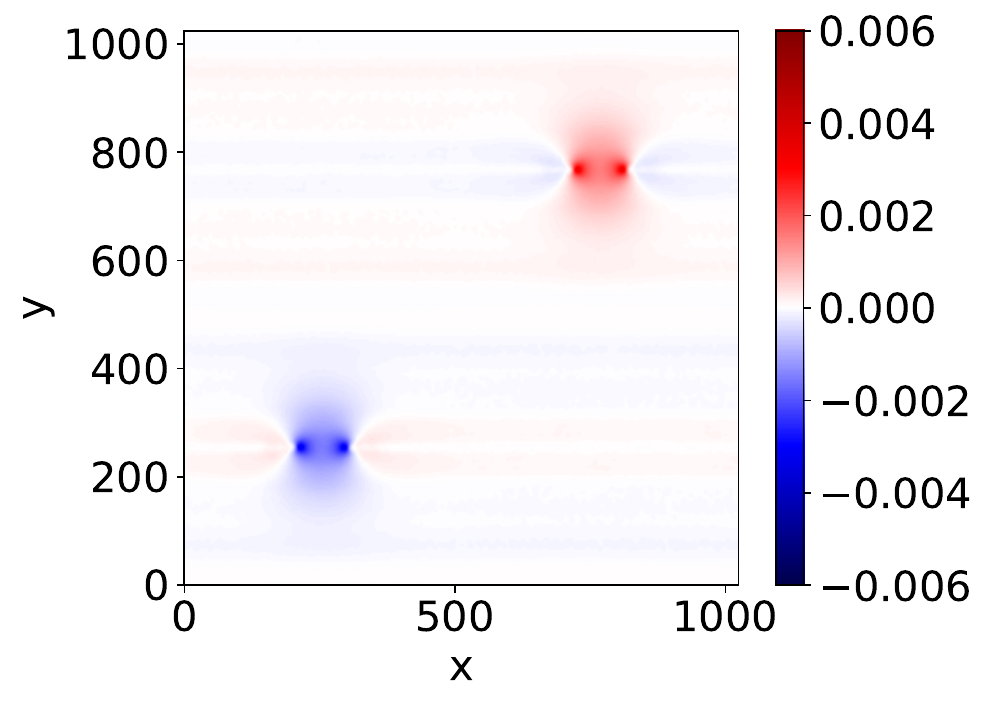}
        \caption{Fault free}
        \label{fig:reconnection_ok}
    \end{subfigure}
    \hfill
    \begin{subfigure}[]{0.47\linewidth}
        \centering
        \includegraphics[width=\linewidth]{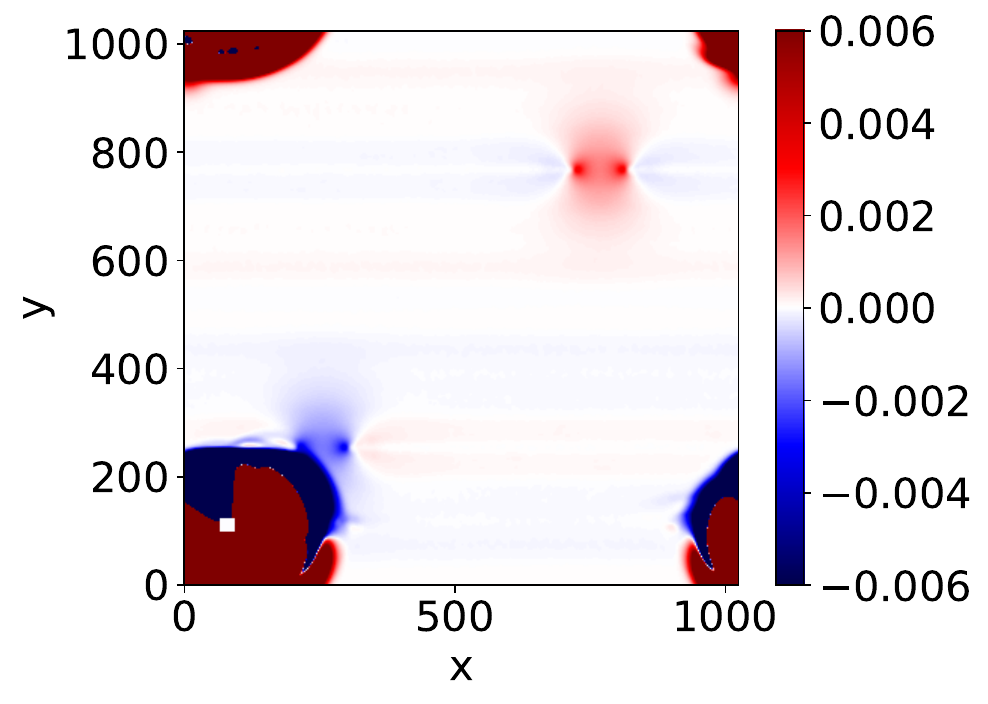}
        \caption{Faulty}
        \label{fig:reconnection_ko}
    \end{subfigure}
    \caption{Comparison between the resulting Electric fields along the z-axis for the executions of the magnetic reconnection experiment in the presence and absence of faults. The same normalisation as in Figure~\ref{fig:weibel} is employed.}
    \label{fig:reconnection}
\end{figure*}

Figures~\ref{fig:reconnection_ok} and~\ref{fig:reconnection_ko} compare the electric field plots along the z-axis at iteration 1980, 180 after the fault. It is easy to see how the fault corrupts a large area, yet the area where the reconnection occurs (top-right quadrant) remains almost untouched. Despite the significant impact on the field values, the fault does not manage to change the measured magnetic reconnection value. Thus, the faulty execution yields a \textbf{100\% accurate result}. 

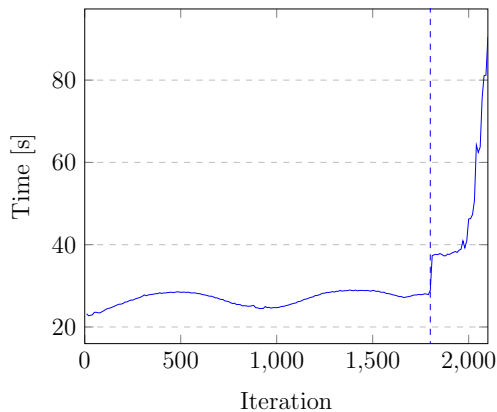
\begin{figure}[t]
    \centering
    \pgfplotsset{
/pgfplots/boxplot legend/.style={
   legend image code/.code={
     \draw [|-|,##1] (0,1.5mm) --
         node[rectangle,minimum size=2mm,draw,fill,fill opacity=1,##1]{}
        (0,5mm);
    }
  }
}

\begin{tikzpicture}[scale=0.6,transform shape]

\begin{axis}[
    title style={at={(0.5,-0.3)},anchor=north},
    xlabel style={at={(0.5,-0.12)},anchor=north},
    xlabel={Iteration},
    ylabel={Time [s]},
    xmin=0,
    xmax=2100,
    legend style={at={(0.5,-0.2)},
    anchor=north,legend columns=-1},
    ymajorgrids=true,
    y grid style=dashed,
]

\draw[dashed,blue] (1800,0) -- (1800,200);

\addplot[color=blue] plot coordinates {};

\addplot [color=blue] table[y=diff,x=time,col sep=comma] {plots/large_timings.csv};

    


\end{axis}

\end{tikzpicture}
    \caption{Time to complete 10 iteration of the algorithm as a function of the iteration for the execution with faults of the magnetic reconnection experiment.}
    \label{fig:reconnection_time}
\end{figure}

The behaviour of the execution time per iteration follows a similar pattern as the previous experiment, as shown in Figure~\ref{fig:reconnection_time}: after a fault, the time needed to perform the movement of the particle raises as we need to change the algorithm, and the corruption of the fields implies that particles are subject to anomalous forces, moving way more than they would have in a fault-free scenario. These circumstances cause an increase in the execution time after fault insurgence, a pattern that becomes even more dramatic in later iterations.

This second test highlighted another circumstance in which the fault did not affect the result: if it involves an area which does not participate in the computation of the metric of interest, its impact is negligible. This statement loses validity the more iterations elapse since fault insurgence and the closer the fault is to the area of interest. Nonetheless, this experiment proved that the data may remain valid for many iterations, as the faulty execution lasted more than 300 iterations after the fault.

These experiments prove that it is possible to extract meaningful values from a faulty execution even without rolling back to the previously saved checkpoint, providing a valuable alternative when only a few iterations remain from the target measurement. While graceful degradation properties may seem an alternative to traditional C/R mechanisms usually employed in stencil applications, they cannot replace them entirely: we showed how graceful degradation properties operate, but we also highlighted that sometimes the execution still produces no meaningful result, as the fault corruption may affect the area of interest too early or too significantly. Additionally, the increase in execution times after a fault implies that rolling back to the previous saved state may be more beneficial than continuing with the current faulty execution, wasting time at each iteration. Nonetheless, we think that combining a traditional C/R solution with the graceful degradation possibility can only bring benefits in applications like iPiC3D, as we can choose whether to recover the previous checkpoint or continue the execution considering the location of the fault, the number of iterations remaining, and the number of iterations since last checkpoint.

\section{Conclusions}\label{sec:conclusion}


In this paper, we considered the current status of fault resilience properties applied to stencil applications, overcame the limitations of the Legio framework and applied these additional functionalities to a real-world stencil application over two actual use cases. With the decoupling between communicators and topologies, we could propagate the information about topologies through fault insurgence and repair, and by leveraging the alternative incomplete topology strategies, we can deal with the absence of processes. Combining Legio with the analysed stencil application required some minor adjustments, mainly due to changing the particle routing algorithm to account for the possibility of missing processes. After the integration, we evaluated the impact of faults on the use cases: despite fault presence, we managed to scavenge the required values from the executions. This result is a remarkable achievement, as it proves the benefits of graceful degradation in application with tight communication between the processes.


Overall, introducing Legio fault resilience functionalities in a stencil application proved non-trivial. In this case, the analysis of the communication patterns employed by the iPiC3D application revealed the criticality of the particle mover algorithm. By pairing the original algorithm with an A* implementation, we overcame the issue but introduced some overhead in the execution. Other stencil applications may not expose this limitation, thus benefitting even more from the additional possibilities coming from Legio.

The graceful degradation solution we propose in this paper does not want to be an all-around alternative to the usual C/R solution, as it does not always work regardless of fault location and timing. Nonetheless, leveraging its benefits may reduce the time to result, as reloading the state can be costly and wastes all the computation that happened after it. Combining C/R and graceful degradation can yield promising results, continuing execution when close to the end of the execution and reloading the previous checkpoint otherwise. Moreover, it could be possible to automatically choose the recovery mechanism depending on fault location and timing, but we leave this interesting research direction for future work.

\section*{Acknowledgements}
\noindent{This work was supported by the Italian Ministry of University and Research (MUR) under the PON/REACT project and partially by the SPACE project, funded by the European Union. 
SPACE has received funding from the European High Performance Computing Joint Undertaking (JU) and Belgium, 
Czech Republic, France, Germany, Greece, Italy, Norway, and Spain under grant agreement No 101093441..
We also acknowledge EuroHPC Joint Undertaking for awarding us access to Karolina at IT4Innovations, Czech Republic (EHPC-DEV-2023D10-018).}

\section*{CRediT authorship contribution statement}
\noindent\textbf{Roberto Rocco:} Conceptualisation, Methodology, Software, Validation, Investigation, Writing - Original Draft, Writing - Review \& Editing, Visualisation.
\textbf{Elisabetta Boella:} Conceptualisation, Validation, Investigation, Resources, Writing - Original Draft, Writing - Review \& Editing.
\textbf{Daniele Gregori:} Resources, Writing - Review \& Editing, Funding acquisition.
\textbf{Gianluca Palermo:} Writing - Review \& Editing, Supervision, Funding acquisition.

\bibliographystyle{elsarticle-num} 
\bibliography{thesis}

\end{document}